\documentclass[prx,twocolumn,superscriptaddress,aps,longbibliography,10pt]{revtex4-2}
\usepackage{natbib}
\usepackage{graphicx}
\usepackage{enumitem}
\DeclareGraphicsExtensions{.pdf,.png,.jpg}
\usepackage{array}
\usepackage{amsmath}
\usepackage{amsthm}
\usepackage{amssymb}    
\usepackage{latexsym}
\usepackage{amsfonts}
\usepackage{mathrsfs}
\usepackage{color}
\usepackage{bbold}
\usepackage{kotex}
\usepackage{float}
\usepackage{makecell}
\usepackage{lineno}
\makeatletter
\let\newfloat\newfloat@ltx
\makeatother
\usepackage{algorithm}
\usepackage{algpseudocode}
\usepackage[english]{babel}
\newtheorem{theorem}{Theorem}
\newtheorem*{corollary}{Corollary}
\newtheorem*{lemma}{Lemma}

\usepackage{subfigure}
\usepackage[colorlinks=true, urlcolor=blue,citecolor=blue,linkcolor=blue]{hyperref}
\usepackage{xcolor}

\begin{document}
\newcommand{\bra}[1]{\left< #1\right|}   
\newcommand{\ket}[1]{\left|#1\right>}
\newcommand{\abs}[1]{\left|#1\right|}
\newcommand{\ave}[1]{\left<#1\right>}
\newcommand{\Tr}{\mbox{Tr}}
\renewcommand{\d}[1]{\ensuremath{\operatorname{d}\!{#1}}}
\renewcommand\qedsymbol{$\blacksquare$}
\newcommand{\argmin}{\arg\!\min}
\newcommand{\argmax}{\arg\!\max}

\title{Operational Quasiprobability in Quantum Thermodynamics: Work Extraction by Coherence and Non-joint Measurability}

\author{Jeongwoo Jae}
\email{jeongwoo.jae@samsung.com}
\affiliation{R\&D center, Samsung SDS, Seoul, 05510, Republic of Korea}

\author{Junghee Ryu}
\affiliation{Center for Quantum Information R\&D, Korea Institute of Science and Technology Information (KISTI),\\ Daejeon 34141, Republic of Korea}
\affiliation{Division of Quantum Information, KISTI School, Korea University of Science and Technology, Daejeon 34141, Republic of Korea}

\author{Hoon Ryu}
\email{elec1020@kumoh.ac.kr}
\affiliation{School of Computer Engineering, Kumoh National Institute of Technology, Gumi, Gyeongsangbuk-do 39177, Republic of Korea}


\begin{abstract}
We employ the operational quasiprobability (OQ) as a work distribution, which reproduces the Jarzynski equality and yields the average work consistent with the classical definition. The OQ distribution can be experimentally implemented through the end-point measurement and the two-point measurement scheme. Using this framework, we demonstrate the explicit contribution of coherence to the fluctuation, the average, and the second moment of work. In a two-level system, we show that non-joint measurability, a generalized notion of measurement incompatibility, can increase the amount of extractable work beyond the classical bound imposed by jointly measurable measurements. We further prove that the real part of Kirkwood-Dirac quasiprobability (KDQ) and the OQ are equivalent in two-level systems, and they are nonnegative for binary unbiased measurements if and only if the measurements are jointly measurable. In a three-level Nitrogen-vacancy center system, the OQ and the KDQ exhibit different amounts of negativities while enabling the same work extraction, implying that the magnitude of negativity is not a faithful indicator of nonclassical work. These results highlight that coherence and non-joint measurability play fundamental roles in the enhancement of work.
\end{abstract}
\maketitle

\section{introduction}
Work is a fundamental quantity in thermodynamics, bridging nonequilibrium processes to equilibrium properties. The Jarzynski equality (JE) provides a remarkable relation: the exponential average of work performed during a nonequilibrium process equals the free energy difference between equilibrium states~\cite{Jarzynski1997}. In classical systems, work can be defined as a stochastic variable along a trajectory. In quantum systems, however, the lack of well-defined trajectories and the back-action of measurement render the definition of work ambiguous~\cite{Talkner2007}. This conflict is formalized in no-go theorems, which state that no quantum probability distribution can simultaneously satisfy the JE and reproduce the classical definition of average work~\cite{Perarnau2017,Lostaglio2023kdq}. To address this limitation, recent studies have employed quasiprobabilities which generalize classical probability distributions by allowing negative or even complex values~\cite{Francica2022,Francica2022most,Gherardini2024}. This approach naturally raises a fundamental question: How are these anomalous values connected to nonclassical features such as coherence and measurement incompatibility in the quantum thermodynamics?

Recently, the Kirkwood–Dirac quasiprobability (KDQ) has attracted attention as a work distribution in quantum systems~\cite{Gherardini2024}. The KDQ is defined in terms of observables and retains its basis representation, enabling an operationally consistent interpretation with quantum measurement statistics. Its usefulness has been demonstrated in a wide range of fields, including quantum metrology~\cite{Arvidsson2020,Noah2022}, many-body physics~\cite{Yunger2017,Yunger2018}, and the fundamentals of quantum physics~\cite{Arvidsson2024}. More recently, negativity in KDQ has been shown to enhance extractable work exceeding the classical bound~\cite{Santiago2024}. The KDQ negativity indicates that a quantum state does not commute with one of measurements considered or the measurements are mutually noncommuting; however, the converse does not necessarily hold~\cite{Lostaglio2023kdq}. This implies that the negativity represents a stricter notion of nonclassicality than noncommutativity~\cite{Arvidsson2021}. Since various forms of nonclassical thermodynamic behavior are linked to the KDQ negativity, it is crucial to clarify the underlying operational principles. In particular, {\em non-joint measurability}---a generalized notion of measurement imcompatibility~\cite{Busch86,Heinosaari2016,Beyer2022,Guhne2023} plays a central role in characterizing nonclassicality such as the quantum steering~\cite{Quintino2014,Uola2015}, the wave-particle duality~\cite{Liu09}, and the uncertainty relation~\cite{Busch2013,Busch2014}. In this context, alternative frameworks, such as the operational quasiprobabiltiy~\cite{Ryu2013,Jae2017}, may provide a direct and operational connection to non-joint measurability. Moreover, the negativity of the operational quasiprobability is an indicator of nonclassicality, associated with phenomena such as entanglement~\cite{Ryu2013}, violation of macrorealism~\cite{Jae2017}, measurement-selection contextuality~\cite{Ryu2019}, and non-joint measurability~\cite{Jae2019}.

In this work, we employ the operational quasiprobability (OQ) as a work distribution which can be experimentally implemented through end-point and two-point measurement. We investigate the thermodynamic properties of the OQ, and show that it reproduces both the JE and the classical definition of average work. For coherent states, the JE derived from the OQ includes a modification term, which captures the contribution of coherence to the fluctuation, the average, and the second moment of work. The work distribution defined by the OQ shows that the non-joint measurability enables nonclassical enhancement of work extraction in a two-dimensional system. This provides an operational identification of non-joint measurability in thermodynamic systems, which is of great importance in the fundamentals of quantum physics and quantum information science~\cite{Guhne2023}. We also show that the OQ is equivalent to the real part of the KDQ, called the Margenau–Hill quasiprobability (MHQ), for arbitrary binary measurements. This equivalence unravels that the MHQ of binary unbiased measurements becomes negative {\em if and only if} the considered measurements are non-jointly measurable. In a three-dimensional Nitrogen-vacancy center system~\cite{Santiago2024}, the OQ and MHQ exhibit different negativities but yield the same extractable work. These results highlight the importance of coherence and non-joint measurability in the enhancement of work.

\section{Quantum thermodynamics using OQ}

\subsection{Work protocol and Two-point measurement}
We define the quantum work based on the following experimental protocol: Consider that a $d$-dimensional quantum state $\hat{\varrho}$ evolves by a time-dependent Hamiltonian
\begin{eqnarray}
\label{eq:Ham}
    \hat{ H}(t) = \sum_{x=0}^{d-1} E_{x}\hat{\Pi}_{x}(t),
\end{eqnarray}
where $\Pi_{x}(t)$ is a projector onto the eigenspace associated with the eigenvalue $E_x$ at time $t$. The Hamiltonian defines the energy observable at each time. We measure the energy of the state using two projective measurements. The measurements performed at time $t_1$ and $t_2$ ($t_1 < t_2$) are defined by $A:=\{\hat{A}_{i}=\hat{\Pi}_i(t_1)\}$ and $B:=\{\hat{B}_{f}=\hat{\Pi}_f(t_2)\}$, respectively, where $i,f\in[d]$ are the measurement outcomes. The time evolution of the quantum state can be described by a completely positive and trace-preserving (CPTP) map $\Phi_H$. We express the measurement at time $t_2$ in the Heisenberg picture as $B^H := \{\hat{B}^H_f = \Phi_H^\dagger(\hat{\Pi}_{f}(t_2))\}$, where $\Phi_H^\dagger$ is a dual map of the CPTP map which is unital. In the following examples, we mainly consider closed systems in which the time evolution of a state is governed by a unitary operator and the work is defined by the measured energy difference.

In this protocol, the typical method to define average work is to use the two-point measurement (TPM) scheme which performs the measurements $A$ and $B$ consecutively~\cite{Talkner2007}. The probability of the TPM scheme can be read as $p^\text{TPM}_{if} = p^{B|A}_{f|i}p^A_i$, where $p^A_i = \Tr(\hat{\varrho}\hat{A}_{i})$ is the probability obtained by the measurement $A$, and $p^{B|A}_{f|i}=\Tr (\hat{A}^{1/2}_{i}\hat{\varrho}\hat{A}^{1/2}_{i}\hat{B}^H_{f})/p^A_i$ is the conditional probability of $B$ given that $A$ has been measured first. Talkner showed that the TPM scheme reproduces the JE for a Gibbs state and arbitrary unitary evolution~\cite{Talkner2007}. However, because the first measurement inevitably disturbs the state $\hat{\varrho}$, the TPM scheme fails to reproduce the average work defined by the energy difference $E_f - E_i$, where $E_f$ and $E_i$ denote the energies measured at time $t_2$ and $t_1$, respectively. This discrepancy arises from the mismatch between the marginal of the TPM probability and the probability of individual measurement;
\begin{eqnarray}
\label{eq:TPMmarginal}
    \sum_{f} p^\text{TPM}_{if} = p^A_i \quad\text{and}\quad \sum_{i} p^\text{TPM}_{if} \neq p^B_f.
\end{eqnarray}
In quantum theory, no positive joint probability distribution can simultaneously satisfy the JE and preserve the marginals of both measurements. Enforcing both properties inevitably results in negative values in the distribution. To circumvent this problem, quasiprobabilities have been employed to define work and its distribution~\cite{Francica2022,Francica2022most,Gherardini2024}.

\begin{figure}
    \centering
    \includegraphics[width=0.97\linewidth]{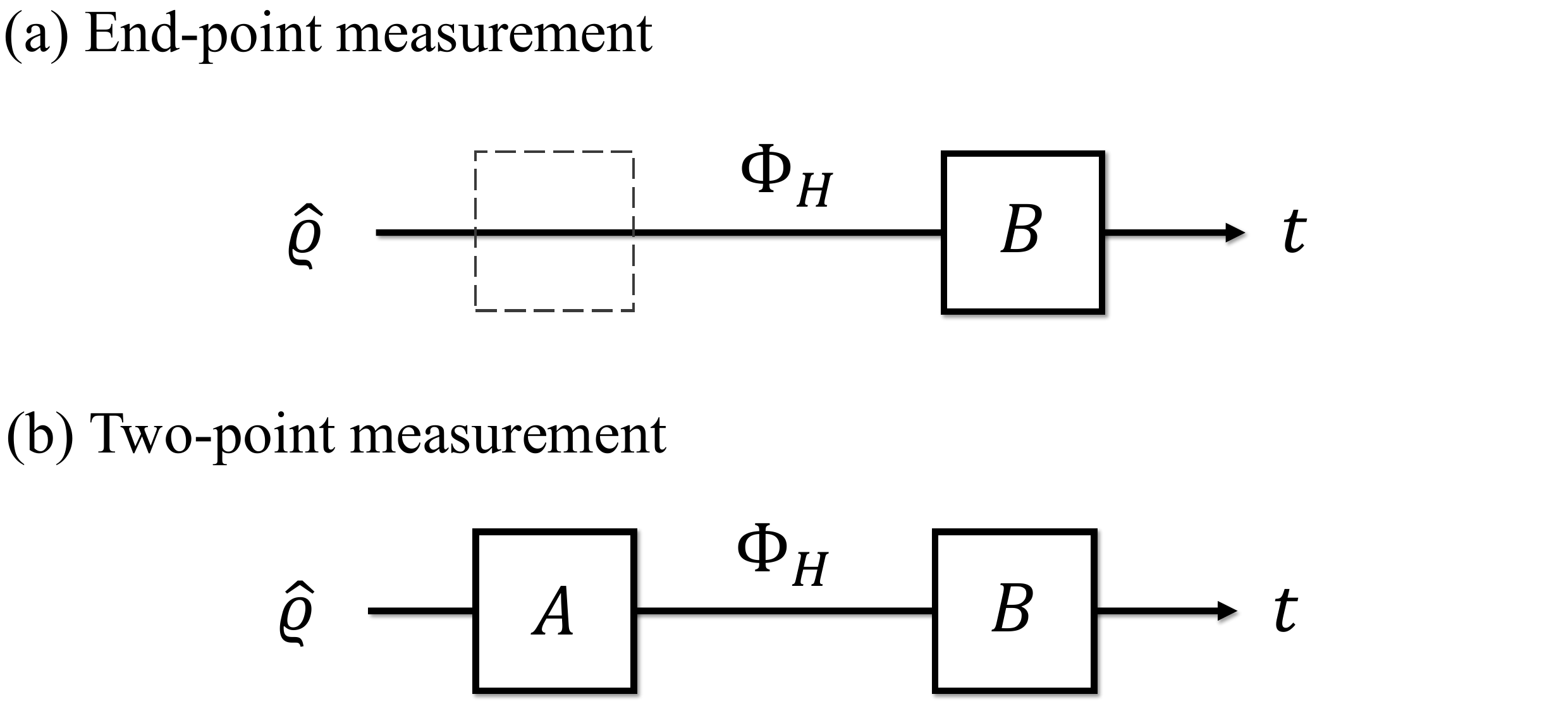}
    \caption{Measurement settings to obtain the operational quasiprobability. (a) End-point measurement (EPM) performs the measurement $B$ at time $t_2$. (b) Two-point measurement (TPM) is a consecutive measurement performing the measurement $A$ and $B$ at time $t_1$ and $t_2$ ($t_1<t_2$), respectively. The input state $\hat{\varrho}$ evolves according to the time-dependent Hamiltonian ${H}(t)$ defined in~\eqref{eq:Ham}. The quantum channel $\Phi_H$ represents the time evolution by the Hamiltonian ${H}(t)$.}
    \label{fig:measurement}
\end{figure}

\subsection{Thermodynamic properties of OQ}

We here consider the OQ as a work distribution. The OQ is defined by the two settings consisting of the measurements $A$ and $B^H$ : The setting (a) in Fig.~\ref{fig:measurement}, called the end-point measurement (EPM), performs the measurement $B$ at time $t_2$. The probability of the EPM is given by $p^\text{EPM}_f = p^B_f=\Tr(\hat{\varrho} \hat{B}^H_{f})$. The setting (b) in Fig.~\ref{fig:measurement} is the TPM scheme. Based on the probabilities of the EPM and the TPM scheme, we have the function of OQ,
\begin{eqnarray}
\label{eq:OQ}
    q^{\text{OQ}}_{if}:=p^\text{TPM}_{if}+\frac{1}{d}\left(p^\text{EPM}_{f} - \sum_{i=0}^{d-1} p^\text{TPM}_{if}\right).
\end{eqnarray}
This function is derived from the inverse Fourier transform of the characteristic functions of the two measurement settings (see Appendix~\ref{sec:prop} for details). Note that the representation of the OQ is not altered depending on a functional form of observables, unlike the quasiprobabilities such as the Wigner function. Experimental verification of the OQ is achieved in optical systems~\cite{Ryu2013}, and the framework is extended to continuous variable systems in Ref.~\cite{Jae2017}.

We identify key properties of the OQ which can be exploited to demonstrate its relevance in quantum thermodynamics:
\begin{enumerate}[label={(T\arabic*)}]

    \item {\em Marginality:} The OQ reproduces the probabilities of the measurements $A$ and $B^H$ as its marginals, i.e., $\sum_f q^\text{OQ}_{if} = p^A_i$ and $\sum_i q^\text{OQ}_{if} = p^B_f$.

    \item {\em TPM reproducibility:} For an input state $\hat{\varrho}=\hat{\varrho}_D$ which is diagonalized in the basis of the first measurement $A$, the OQ becomes the probability of the TPM scheme, i.e., $q^\text{OQ}_{if}=p^\text{TPM}_{if}$.

    \item {\em Convex linearity:} For an input state $\hat{\varrho}=\sum_k p_k\hat{\varrho}_k$ where $\sum_k p_k=1$ and $p_k \ge 0 $ $\forall k$, the OQ is a linear functional such that $q^\text{OQ}_{if}(\hat{\varrho})=\sum_k p_k q^\text{OQ}_{if}(\hat{\varrho}_k)$.
    
\end{enumerate}
The properties (T$2$) and (T$3$) can be easily obtained by the form of the OQ function~\eqref{eq:OQ}. The property (T$2$) is obtained by the fact that, as a diagonal state commutes with the measurement $A$ such that $[\hat{\varrho}_D, \hat{A}_i]=0$ $\forall i$, the probability of the TPM can be read as $p^\text{TPM}_{if}=\Tr(\hat{\varrho}_D \hat{A}_i \hat{B}^H_f)$, and the terms in the parenthesis of~\eqref{eq:OQ} are vanished since $\sum_i p^\text{TPM}_{if} = p^\text{EPM}_f$. Note that a more general form of the no-go theorem can be formulated with the conditions (T$1$) and (T$3$)~\cite{Lostaglio2023kdq}. We investigate the thermodynamic significance of the OQ as a work distribution based on these properties.

The work is defined by the average difference between energies measured at different times. By the marginality (T$1$), the expectation of the energy difference over the OQ, $\langle{w}\rangle_\text{OQ}:=\langle{E_f-E_i}\rangle_\text{OQ}$, coincides with the difference between the average energies obtained by the measurements $A$ and $B$ separately as
\begin{eqnarray}
\label{eq:avg}
    \langle{w}\rangle_\text{OQ} = \sum_{i,f} q^\text{OQ}_{if}\left( E_{f} - E_{i} \right) = \langle{E}_f\rangle_B - \langle{E}_i\rangle_A,
\end{eqnarray}
where the subscript $A (B)$ implies that the expectation is taken over the probability obtained by the measurement $A(B)$.

The condition (T$2$) is important to derive the JE and subsequently the fluctuation theorem. For a Gibbs state $\hat{\varrho}_G$, JE holds for arbitrary unitary evolution~\cite{Talkner2007} as
\begin{eqnarray}
    \langle e^{-\beta w} \rangle_\text{OQ} = \sum_{i,f}q^\text{OQ}_{if} e^{-\beta (E_f-E_i)} = e^{-\beta \Delta F},
\end{eqnarray}
where $\Delta F$ is the free energy difference given by the ratio of the partition functions of the equilibrium states at time $t_2$ and $t_1$, and $\beta=(k_BT)^{-1}$ is the inverse temperature with the Boltzmann constant $k_B$. Applying Jensen's inequality, JE implies $\langle w \rangle \ge \Delta F$ by the second law of thermodynamics.

For a coherent input state $\hat{\varrho} = \hat{\varrho}_D + \hat{\varrho}_\text{off}$ with $\hat{\varrho}_D$ the diagonal and $\hat{\varrho}_\text{off}$ the off-diagonal components in the basis of the measurement $A$, JE in the TPM scheme does not retain its conventional form. Instead, it becomes $\langle e^{-\beta w} \rangle_\text{TPM} = e^{-\beta \Delta F}\Gamma_\text{TPM}$, where $\Gamma_\text{TPM} = \Tr[\hat{\varrho}_{G,i}^{-1}\hat{\varrho}_D \Phi^\dagger_H(\hat{\varrho}_{G,f})]$~\cite{Gherardini2024}. The additional term $\Gamma_\text{TPM}$ implies that the work fluctuation derived from the TPM scheme only represents the effect induced by the incoherent elements of the initial state, as the back-action of the measurement $A$ removes the coherence of the initial state.

For coherent states, the JE of the OQ becomes
\begin{eqnarray}
\label{eq:OQmod}
    \langle e^{-\beta w} \rangle_\text{OQ} = e^{-\beta \Delta F}\Gamma_\text{OQ}.
\end{eqnarray}
As the statistics of the EPM and the marginal probability of TPM are involved, the additional term is given by
\begin{eqnarray}
\label{eq:OQmodi}
    \Gamma_\text{OQ} = \Gamma_\text{TPM} + \frac{1}{d}\Tr\left(\hat{\varrho}^{-1}_{G,i}\right)\Tr\left[\hat{\varrho}_\text{off} \Phi^\dagger_H\left(\hat{\varrho}_{G,f}\right)\right].
\end{eqnarray}
See Appendix~\ref{sec:gammaOQ} for details. This implies that $\Gamma_\text{OQ}=\Gamma_\text{TPM}$ when the input state has no coherence $\hat{\varrho}=\hat{\varrho}_D$. Unlike the TPM scheme, the additional term $\Gamma_\text{OQ}$ considers the overlap between the coherent state at the initial time and the Gibbs state at the final time. This shows that the OQ formalism for the fluctuation relations captures how the initial coherence influences the final equilibrium state, relative to the energy landscape of the initial equilibrium state. Note that the additional term for the KDQ is given by $\Gamma_\text{KDQ}= \Gamma_\text{TPM} + \Tr[\hat{\varrho}_{G,i}^{-1}\hat{\varrho}_\text{off}\Phi^\dagger_H(\hat{\varrho}_{G,f})]$~\cite{Gherardini2024}.

The $n$-th moment of work is given by the derivative of the generating function as $\langle w^n \rangle_\text{OQ} = \Delta F^n +(-1)^n{\partial^n}\Gamma_\text{OQ}/{\partial^n \beta} \vert_{\beta=0}$. Thus, the average of work becomes
\begin{eqnarray}
\label{eq:wcoh}
    \langle w \rangle_\text{OQ} = \langle w \rangle_\text{TPM} + \Tr\left(\hat{\varrho}_\text{off} \hat{H}^H_f\right),
\end{eqnarray}
where $\langle w \rangle_\text{TPM}$ represents the average work obtained from the incoherent state $\hat{\varrho}_D =  \sum_i \hat{A}^{1/2}_i\hat{\varrho}\hat{A}^{1/2}_i$. This result shows that, for incoherent state, the work of the OQ is equivalent to that of the TPM scheme by the condition (T$2$). For the work difference between the OQ and the TPM scheme $\delta w := \langle w \rangle_\text{OQ} - \langle w \rangle_\text{TPM}$, we can obtain a relation similar to the second law of thermodynamics:
\begin{eqnarray}
\label{eq:Fco}
    \delta w = \delta F + k_BT{\cal C}_\text{rel}\left(\hat{\varrho}\right),
\end{eqnarray}
where $\delta F = F\left(\hat{\varrho}\right)-F\left(\hat{\varrho}_D\right)$, $ F(\hat{\sigma})=\Tr(\hat{\sigma}\hat{H}_f^H) - kTS(\hat{\sigma})$ is the Helmholtz free energy, ${\cal C}_\text{rel}(\hat{\varrho})=S(\hat{\varrho}_D)-S(\hat{\varrho})$~\cite{Baumgratz2014} is the relative entropy of coherence measure, and $S(\hat{\varrho})=-\Tr(\hat{\varrho}\log\hat{\varrho})$ is the Von Neumann entropy. Thus, the work difference can be decomposed into (i) the free-energy difference between the coherent state and its incoherent counterpart and (ii) the amount of coherence. Since ${\cal C}_\text{rel}\ge 0$, the nonclassical contribution to the OQ work is bounded from below by the free-energy difference $\delta w \ge \delta F$. The equality holds {\em if and only if} $\hat{\varrho}$ is an incoherent state.

Furthermore, the bound of~\eqref{eq:wcoh} can be determined by the initial coherence of the input state as
\begin{eqnarray}
\label{eq:l1bound}
    \abs{\delta w} \le {\cal C}_{l_1}(\hat{\varrho})\left\Vert \hat{H}^H_{f,\text{off}}\right\Vert_{\infty},
\end{eqnarray}
where ${\cal C}_{l_1}$ is the coherence measure of the (entrywise) $1$-norm~\cite{Baumgratz2014} and $\Vert \hat{H}^H_{f,\text{off}}\Vert_\infty = \max_{i\neq j}|\hat{H}^H_{f,ij}|$ is the (entrywise) $l_\infty$-norm and the the off-diagonal terms are defined in the basis of $\hat{\varrho}$. To obtain this relation, we apply H{\"o}lder’s inequality to~\eqref{eq:wcoh}. The equality holds if and only if the absolute value of $\hat{H}^H_{f,\text{off}}$ attains the same maximal value at every index where $\hat{\varrho}_\text{off}$ is nonzero. These results highlight the contribution of coherence to the average work, which cannot be captured within the TPM scheme.

The second moment is of great importance in the operational determination of the fluctuation of systems. In many cases, one usually considers the symmetric energy landscape such that $\Tr \hat{H}_i = \sum_i E_i=0$. Under this condition, the second moment of work of the OQ is lower bounded as
\begin{eqnarray}
\label{eq:w2bound}
    \langle w^2\rangle_\text{OQ} \ge \Tr\left[ \hat{\varrho}_\text{off}\left(\hat{H}^H_f\right)^2\right]. 
\end{eqnarray}
The proof is presented in Appendix.~\ref{sec:boundw2}. This relation shows that, unlike the quasiprobabilities listed in Ref.~\cite{Francica2022most}, the second moment of work in the OQ can be negative, but it is lower bounded by the coherent part of the initial state.

We have shown that the OQ can serve as a work distribution satisfying the conditions (T$1$)--(T$3$) and that the work difference between the OQ and the TPM scheme is originated from the coherence of initial state. In Appendix.~\ref{sec:traceCoh}, we also show that the difference between the OQ and the TPM probability represents the trace norm of coherence parts of input state. In the following sections, we investigate the enhancement of work extraction enabled by non-joint measurability and compare the OQ to the KDQ.

\begin{figure*}[t!]
    \centering
    \includegraphics[width=\linewidth]{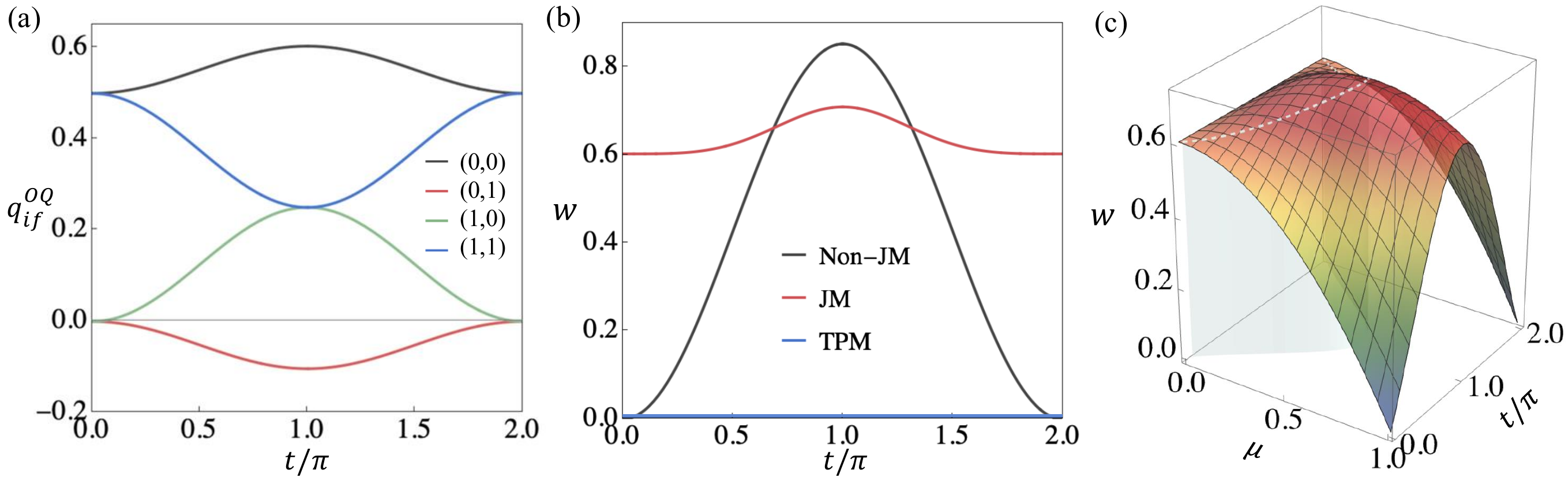}
    \caption{The work extraction based on the operational quasiprobability (OQ) in a single-qubit system. (a) Operational quasiprobability $q^\text{OQ}_{if}$, and the tuple $(i,f)$ denotes the outcomes. The OQ is negative for the excitation process corresponding to the outcome $(0,1)$. (b) The black solid line represents the amount of extractable work $w$ by non-joint measurability (non-JM) and the red solid line represents the classical bound given by the jointly measurability (JM). The black solid line is obtained by using the sharp measurement where $\mu=1$. The classical bound is obtained by optimizing the state-independent bound~\eqref{eq:bound} over the measurement sharpness $\mu$. The amount of work obtained by the two-point measurement (TPM) scheme is zero (blue solid line). (c) The landscape of classical upper bound of the extractable work by jointly measurable measurements is illustrated. The white dashed line represents the JM bound in (b).}
    \label{fig:result}
\end{figure*}

\section{Enhanced extractable work by non-joint measurability}

Our results show that non-joint measurability in the OQ framework gives rise to extractable work that exceeds the classical bound imposed by {\em joint measurability}, thereby revealing a nonclassical enhancement. To this end, we extend the work protocol by incorporating generalized measurements~\cite{Beyer2022}, represented by positive operator-valued measures (POVMs). The POVMs performed at time $t_1$ and $t_2$ ($t_1 < t_2$) are defined by $A:=\{\hat{A}_{i}\}$ and $B:=\{\hat{B}_{f}\}$, satisfying $\sum_i\hat{A}_i=I$ and $\sum_f \hat{B}_f=I$ with $i,f\in[d]$. Specifically, we consider that the POVMs are defined over the projectors of the Hamiltonian of system, and they can be represented by unsharp energy measurements as $\hat{A}_i=\sum_m a^i_m\hat{\Pi}_m(t_1)$ and $\sum_{n}b_n^f\hat{\Pi}_n(t_2)$ with $\sum_i a^i_m = 1$ and $\sum_f b_n^f = 1$ for $a^i_m,b^f_n\in[0,1]$. Each element of these measurements becomes sharp when $a^i_m = \delta_{m'm}$ and $b^j_n=\delta_{n'n}$. The measurements considered in this section are POVMs $A=\{\hat{A}_i\}$ and $B^H=\{\Phi^H(\hat{B}_f)\}$ in the Heisenberg picture.

These two POVMs are jointly measurable when there exists a joint POVM, $J=\{\hat{J}_{if}\}$, that can reproduce the two measurements as its marginals, i.e.,
\begin{eqnarray}
    \sum_{f} \hat{J}_{if} = \hat{A}_i \quad \text{and} \quad \sum_{i} \hat{J}_{if} = \hat{B}^H_f,~\forall i,f.
\end{eqnarray}
If the joint measurement $J$ does not exist, the measurements are called non-jointly measurable~\cite{Busch86,Heinosaari2016,Beyer2022,Guhne2023}.

To demonstrate the work enhancement by non-joint measurability, we use the fact that the sign of the OQ depends on the input state $\hat{\varrho}$ and joint measurability between the two measurements used~\cite{Jae2019}. More specifically, we consider binary unbiased measurements which are smeared versions of projective measurements defined in the two-dimensional space ${\cal  H}_{2}$, $\hat{A}_{i} = \mu \Pi_{i}(t_1) + (1-\mu){I}/2$ and $\hat{B}^H_{f} = \mu \Pi_{f}(t_2) + (1-\mu){I}/{2}$, where $\mu$ determines the sharpness of measurements~\cite{Heinosaari2015}. The joint measurability of these POVMs is equivalent to the positivity of the OQ as stated in the following Lemma:

\begin{lemma}
    The operational quasiprobability is positive semidefinite for all two-dimensional quantum states $\hat{\varrho}$ if and only if the binary unbiased measurements $A$ and $B^H$ are jointly measurable (JM), i.e.,
    \begin{eqnarray}
        q^\text{OQ} \ge 0,~\forall\hat{\varrho}\in{\cal  H}_2\quad\Longleftrightarrow\quad A~\text{and}~B^H~\text{are JM}. \nonumber
    \end{eqnarray}
\end{lemma}

The proof of Lemma is presented in Appendix~\ref{sec:lemma} and see Ref.~\cite{Jae2019} for detailed discussions.

We now derive the classical bound of extractable work based on the OQ distribution. The extractable work is defined as the energy remaining at the end of the protocol. So, the requirement for work extraction is
\begin{eqnarray}
    \langle w \rangle = \langle {H}(t_2) \rangle - \langle {H}(t_1)\rangle <0.
\end{eqnarray}
Depending on the contributions to the amount of extractable work, the dynamics of energy can be categorized into two processes; the excitation process, $E_f - E_i >0$, and the de-excitation process, $E_f - E_i \le 0$. For a classical probability, say $p_{if}$, the excitation process reduces the amount of extractable work, and the extractable work can be maximum when $p_{if}=0$ for the excitation processes and $p_{if}>0$ for the de-excitation processes. Thus, the amount of extractable work by the classical distribution is upper bounded as
\begin{eqnarray}
    {\cal W}_\text{cl} = -\langle w \rangle_p \le -\sum_{E_i \ge E_f}p_{if}(E_f - E_i).
\end{eqnarray}
On the other hand, for a quasiprobability $q_{if}$, the amount of work defined by
\begin{eqnarray}
    {\cal W}_q = -\langle w \rangle_q = -\sum_{i,f}q_{if}(E_f - E_i)
\end{eqnarray}
can be further increased by the negative values associated with excitation processes~\cite{Santiago2024}. In the following example, we show that the work derived by the OQ can be larger than those of the classical probability, and the increase of extractable work is induced by the non-joint measurability.

Let us assume the Hamiltonian of a qubit to be
\begin{eqnarray}
    \hat{H}(t)= \frac{1}{2}\left[\Omega\left(\cos(\delta t)\hat{\sigma}_x + \sin(\delta t)\hat{\sigma}_y \right) + \delta \hat{\sigma}_z\right],
\end{eqnarray}
which corresponds to a two-level system subjected to a magnetic field rotating around the $z$-axis. In the rotating frame, the effective Hamiltonian governing the dynamics of the qubit becomes time-independent, i.e., $\hat{H}_\text{eff}(t) = {\Omega}\hat{\sigma}_x/2$. Thus, the evolution operator of the system is given by $\hat{U}= e^{-j\delta \hat{\sigma}_z t /2}e^{-j\Omega \hat{\sigma}_x t /2}$ for $j^2=-1$. The Hamiltonian and its spectrums are given by $\hat{H}(t) = \sum_{x=0}^1 E_x \hat{\Pi}_x(t)$, where $\Delta=\sqrt{\delta^2 + \Omega^2}$, $E_x = \gamma^x\Delta/2$, $\hat{\Pi}_x=(I +\gamma^x \hat{H}(t)/\Delta)/2$, and $\gamma=-1$. We consider the work extracted by the measurements performed at time $t_1 =0$ and $t_2 =t$. For measuring energy, we consider unbiased measurements defined by $\hat{M}_x(t) = \mu \hat{\Pi}_x(t) + (1-\mu){I}/{2}$. We assume that the initial state has coherence in the eigenbases of $\hat{H}(0)$ as
\begin{eqnarray}
    \hat{\varrho} = \begin{pmatrix}
    p & c \\
    c & 1-p
    \end{pmatrix},
\end{eqnarray}
where $0\le p \le 1$ and $c$ is set to a real number for simplicity.

For the Bloch vectors of the input state, $\vec{r}$, and the measurement, $\vec{v}$, the OQ can be read
\begin{eqnarray}
\label{eq:OQ2d}
    q^\text{OQ}_{if} =\frac{1}{4}\left[1+\gamma^{x_i+x_f}\vec{v}_i\cdot\vec{v}_f+\left(\gamma^{x_i}\vec{v}_i+\gamma^{x_f}\vec{v}_f\right)\cdot\vec{r}\right].
\end{eqnarray}
By Lemma, the $q^\text{OQ}_{if} \ge 0$ $\forall i,f$ {\em if and only if} the measurements $M_f$ and $M_i$ are jointly measurable. More specifically, the OQ is positive semidefinite {\em if and only if } $1\pm \vec{v}_i\cdot\vec{v}_f-\Vert\vec{v}_i\pm\vec{v}_f\Vert \ge 0$, and this condition holds {\em if and only if} Busch's criterion is satisfied~\cite{Busch86}; $\Vert \vec{v}_i+\vec{v}_f\Vert + \Vert \vec{v}_i-\vec{v}_f \Vert \le 2$. (See Appendix~\ref{sec:lemma} for details.)

For the positive OQ, we obtain a state-independent upper bound of the extractable work as
\begin{eqnarray}
\label{eq:bound}
    {\cal W}_\text{cl} &\le& -\sum_{E_i\ge E_f}q^\text{OQ}_{if}(E_f - E_i) \nonumber\\
    &\le& \frac{\Delta}{4}\left(1-\vec{v}_i\cdot\vec{v}_f +\left\Vert\vec{v}_i-\vec{v}_f\right\Vert\right),
\end{eqnarray}
where the inequality holds for all quantum states in the two-dimensional Hilbert space, $\hat{\varrho}\in {\cal H}_2$. For the positive distribution, extractable work is maximized when $q^\text{OQ}_{ij}>0$ for the de-excitation process of $E_f < E_i$. This case corresponds to the outcomes $x_i=1$ and $x_f=0$ and the respective amount of work is $-(E_0 - E_1)q^\text{OQ}_{10}=-\Delta q^\text{OQ}_{10}$, where $q^\text{OQ}_{10}$ is given by~\eqref{eq:OQ2d}. This value is maximized when the Bloch vector of the input state becomes $\vec{r}_{\max}=-(\vec{v}_i-\vec{v}_f)/\Vert\vec{v}_i-\vec{v}_f\Vert$. With $\vec{r}_{\max}$, the extractable work is given by the upper bound in~\eqref{eq:bound}.

We say that the nonclassical work extraction appears if
\begin{eqnarray}
    \max {\cal W}_\text{cl} < {\cal W}_\text{q}.
\end{eqnarray}
In Fig.~\ref{fig:result}, we set the parameters of the system to $p=1/2$, $c=1/2$, $\delta = (\sqrt{2}+1)\Omega$, and $\Omega=1$. The sharpnesses of the measurements are assumed to be the same as $\mu$, i.e., $\Vert\vec{v}_i\Vert=\Vert\vec{v}_f\Vert=\mu$. In these settings, the values of the OQ are shown in Fig.~\ref{fig:result} (a). The quasiprobability associated with the excitation process $q^\text{OQ}_{01}$ has negative values and it contributes to the increase in extractable work. Fig.~\ref{fig:result} (b) shows that the non-jointly measurable measurements which are sharp ($\mu=1$) enable the increase in the extractable work beyond the classical bound of the joint measurability (JM). To obtain the JM bound, we further optimize the state-independent bound~\eqref{eq:bound} over the sharpness. So, the increased work extraction by non-joint measurability appears when
\begin{eqnarray}
\label{eq:classbound}
    \max_{\mu}{\cal W}_\text{cl}=\max_{\mu}\frac{\Delta}{4}\left(1-\vec{v}_i\cdot\vec{v}_f +\left\Vert\vec{v}_i-\vec{v}_f\right\Vert\right) < {\cal W}_q.
\end{eqnarray}
Fig~\ref{fig:result} (c) presents the landscape of the upper bound~\eqref{eq:bound} and the maximum value over the sharpness $\mu$ is represented as the white dashed line. These results can be summarized as
\begin{theorem}
    If the amount of extractable work increases beyond the classical bound, $\max_{\mu}{\cal W}_\text{cl}$, the binary unbiased measurements $A$ and $B^H$ are non-jointly measurable.
\end{theorem}
The proof of Theorem $1$ is presented in Appendix.~\ref{sec:non-JM}. Based on Theorem $1$, our method provides an operational method to verify non-JM in the work extraction protocol.

\section{Comparison of OQ and KDQ}

We have shown that the OQ can be used to define quantum work and enable the extractable work to be increased by non-joint measurability in a two-dimensional system. In this section, we compare the OQ with the real part of KDQ, called Margenau-Hill quasiprobability (MHQ). We show that the OQ and the MHQ are equivalent for two-dimensional systems, and this equivalence implies that the joint measurability determines the positivity of MHQ consisting of two binary unbiased measurements. We also consider a three-dimensional system and show that the equivalence does not hold, but they yield the same amount of extractable work.

The KDQ for the two POVMs $A$ and $B^H$ can be defined by
\begin{eqnarray}
    q^\text{KDQ}_{if} := \Tr\left(\varrho \hat{A}_i \hat{B}^H_f\right).
\end{eqnarray}
Several generalizations of KDQ have been proposed~\cite{arvidsson2024properties}. Since this function can in general take imaginary values, its real part is defined as the MHQ and it reads
\begin{eqnarray}
    q^\text{MHQ}_{if} = \text{Re}\left[ q^\text{KDQ}_{if}\right].
\end{eqnarray}
A method to reconstruct the MHQ is to use nonselective measurement scheme~\cite{Johansen2007} in which a state after the measurement at time $t_1$ becomes
\begin{eqnarray}
    \hat{\varrho}_\text{NS,$i$} = \hat{A}^{1/2}_i \hat{\varrho} \hat{A}^{1/2}_i + \hat{A}^{1/2}_{i,C} \hat{\varrho} \hat{A}^{1/2}_{i,C},
\end{eqnarray}
where $\hat{A}_{i,C} = I - \hat{A}_i$ and NS stands for ``nonselective". In a selective scheme, the post-measurement state reduces to an eigenstate of measurement outcome. The reconstruction of MHQ requires the three measurement settings~\cite{Lostaglio2023kdq}: EPM, TPM, and weak-TPM (wTPM) schemes, where the probability of wTPM is given by $p^\text{wTPM}_{if} = \Tr( \hat{\varrho}_\text{NS,$i$}\hat{B}^H_f)$. Combining these probabilities, we have the MHQ function:
\begin{eqnarray}
\label{eq:MHQ_exp}
    q_{if}^\text{MHQ} = p^\text{TPM}_{if} + \frac{1}{2}\bigg(p^\text{B}_{f} - p^\text{wTPM}_{if}\bigg).
\end{eqnarray}
The MHQ is positive semidefinite if the observables considered commute with each other or with an input state~\cite{Lostaglio2023kdq}. This representation will be used to show the equivalence with the OQ.

\begin{figure*}[t!]
    \centering
    \includegraphics[width=\linewidth]{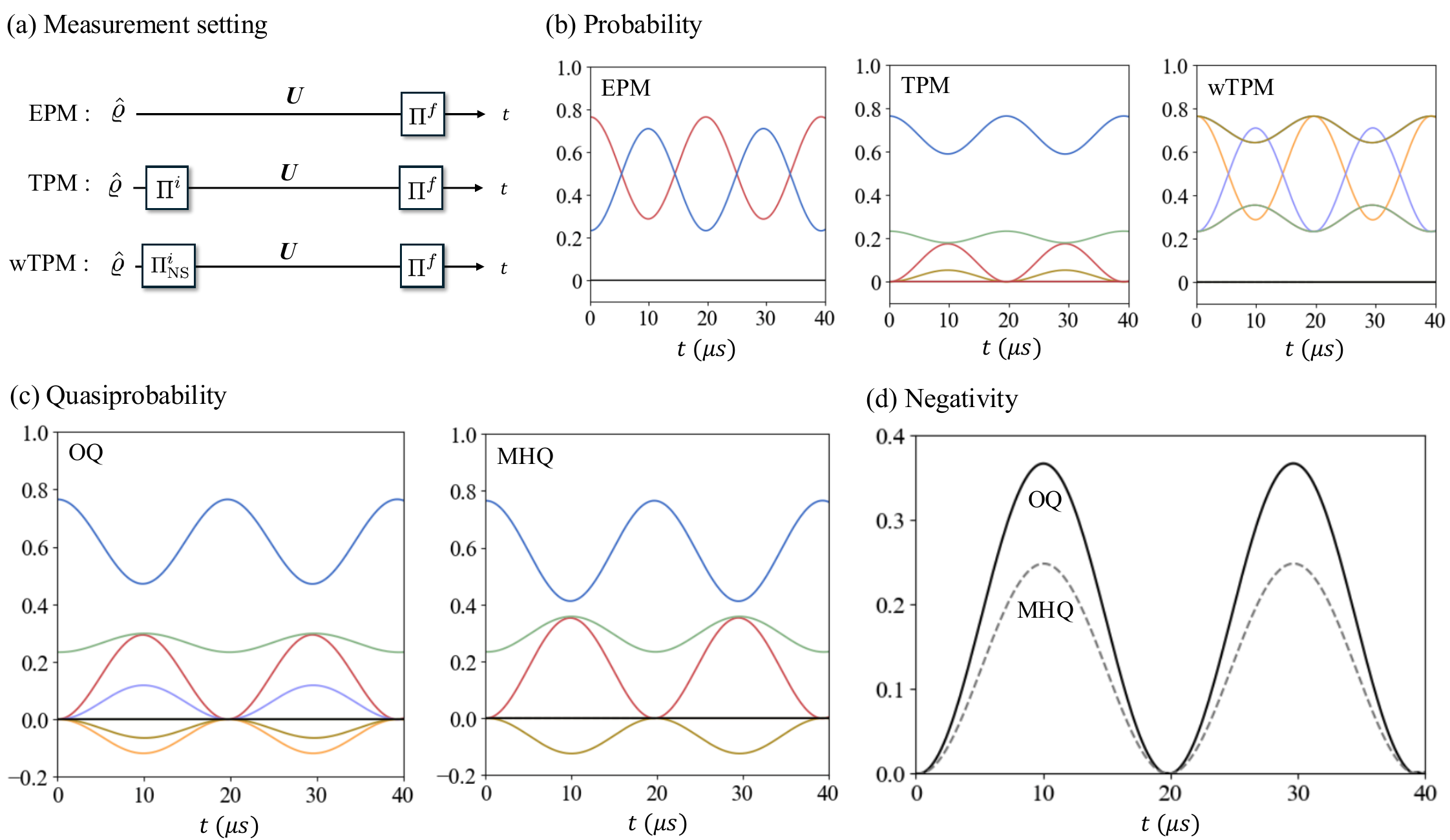}
    \caption{The OQ and MHQ obtained by the three-dimensional system of Nitrogen-vacancy (NV) center in diamond. (a) The OQ is constructed by the EPM and TPM scheme, and the weak-TPM scheme is additionally considered to construct MHQ. (b) shows the probabilities obtained by the measurement settings in (a). (c) The OQ and the MHQ exhibit negative values, but they do not coincide. The OQ negativity shows the visibility higher than those of MHQ.}
    \label{fig:result3d}
\end{figure*}

\subsection{Two-dimensional case}

The equivalence of OQ and MHQ is stated as

\begin{theorem}
    For any binary measurements $A$ defined over the two-dimensional Hilbert space, the OQ and the MHQ are equivalent.
\end{theorem}

The proof of Theorem $2$ is shown in Appendix~\ref{sec:equiv}. The proof is based on the fact that the statistics from the nonselective post-measurement state can be the same with the marginal probability of the TPM scheme for binary measurements, i.e,
\begin{eqnarray}
    p_{if}^\text{wTPM} = \sum_i p_{if}^\text{TPM},
\end{eqnarray}
so the MHQ function~\eqref{eq:MHQ_exp} becomes the OQ function~\eqref{eq:OQ}. For projective measurements, the equivalence between the OQ and the MHQ can be seen in Ref.~\cite{Johansen2007}. We extend this previous result to a general binary measurement which can be biased~\cite{Sixia10}.

Lemma and Theorem $2$ imply that, for the unbiased measurements, their joint measurability is a necessary and sufficient condition for the non-negative MHQ:

\begin{corollary}
    The MHQ is positive semidefinite for all two-dimensional quantum states if and only if the unbiased measurements $A$ and $B^H$ are jointly measurable.
\end{corollary}

It has been known that the negativity of the KDQ can occur only if an initial state noncommutes with one of measurements considered, or if there exists a pair of mutually noncommuting measurements. Corollary reveals that the non-joint measurability is a necessary and sufficient condition for the negative MHQ. Furthermore, this result implies the potential connection between the non-joint measurability and the weak value induced by the negativity of KDQ~\cite{Lostaglio2023kdq,Dressel2014,Wagner2023}, which requires further clarification in future research. Also, since the OQ requires simple measurement settings, it can be an alternative to the MHQ in experiments with two-level systems.

\subsection{Three-dimensional case}

We consider a three-level system, which is a spin-triplet state considered in the experiment of Nitrogen-vacancy (NV) center in diamond~\cite{Santiago2024}. For the $i$-th eigenstate $\ket{i}$ of Hamiltonian, the experiment considered a bichromatic microwave field that resonates with transitions (i) $\ket{0}\rightarrow \ket{-1}$ and (ii) $\ket{0} \rightarrow \ket{1}$. Specifically, in the rotating wave frame of the microwave exerted on the NV center, the Hamiltonian of the system is given by
\begin{eqnarray}
\label{eq:H3dim}
    \hat{H}(t)&=&\Omega_1\left[\hat{S}_{x1}\cos(\phi_1 t)+\hat{S}_{y1}\sin(\phi_1 t)\right] \nonumber\\
    &&+\Omega_2\left[\hat{S}_{x2}\cos(\phi_1 t)-\hat{S}_{y2}\sin(\phi_1 t)\right]
\end{eqnarray}
where $\Omega_1$ and $\Omega_2$ are the Rabi frequencies of the transitions (i) and (ii), respectively, and $\hat{S}$s are the Gell-Mann matrices;
\begin{eqnarray}
    \hat{S}_{x1}&=&\frac{1}{\sqrt{2}}\begin{pmatrix}
        0 & 1 & 0 \\
        1 & 0 & 0 \\
        0 & 0 & 0
    \end{pmatrix},
    ~\hat{S}_{y1}=\frac{1}{\sqrt{2}}\begin{pmatrix}
        0&-i&0\\
        i&0&0\\
        0&0&0
    \end{pmatrix},\nonumber\\
    \hat{S}_{x2}&=&\frac{1}{\sqrt{2}}\begin{pmatrix}
        0 & 0 & 0\\
        0 & 0 & 1\\
        0 & 1 & 0
    \end{pmatrix},
    ~\hat{S}_{y2}=\frac{1}{\sqrt{2}}\begin{pmatrix}
        0 & 0 & 0\\
        0 & 0 & -i\\
        0 & i & 0
    \end{pmatrix}. \nonumber
\end{eqnarray}
The time-independent Hamiltonian in a rotating frame becomes $\hat{H}_\text{eff} = \Omega_1\hat{S}_{x1} - \phi_1\hat{S}_{z1}+\Omega_2\hat{S}_{x2}+\phi_2\hat{S}_{z2}$, where $\hat{S}_{z1}=|1\rangle\langle1|$ and $\hat{S}_{z2}=-|-1\rangle\langle-1|$. The unitary operator of the system is given by
\begin{eqnarray}
    \hat{U}(t) = \exp(-jt\phi_1\hat{S}_{z1})\exp(jt\phi_2\hat{S}_{z2})\exp(-jt\hat{H}_\text{eff}). \nonumber
\end{eqnarray}
where $j^2=-1$.

Fig.~\ref{fig:result3d} (a) shows the measurement settings to construct the quasiprobabilities. The initial state is prepared by a pure state, which minimizes the value of MHQ associated with the excitation process from $\ket{-1}$ to $\ket{1}$~\cite{Santiago2024}, $\ket{\psi}=\sum_i\sqrt{p_i}\gamma^{2\pi j a_i}\ket{i}$, with $p_i=0.7654$, $0.0009$, $0.2338$ and $a_i=0.0073$, $0.2787$, $0.0002$ for $i=1,0,-1$, respectively. We use projective measurements defined by the eigenvector of the Hamiltonian. The Hamiltonian parameters in~\eqref{eq:H3dim} are set to $\Omega_1=\Omega_2=4.4\pi$ MHz and $\phi=1.09\Omega$. In this resonant condition, $\ket{0}$ becomes a dark state by a stimulated Raman adiabatic passage (STIRAP)~\cite{Vitanov2017}, and this effect is shown by the zero value of the EPM probability represented by the black solid line of Fig.~\ref{fig:result3d} (b). Fig.~\ref{fig:result3d} (c) shows the OQ and the MHQ.

We define the negativity of the quasiprobabilities as
\begin{eqnarray}
    {\cal N}\left[q\right] := \sum_{i,f}\abs{q_{if}} - 1.
\end{eqnarray}
(Note that, in Appendix~\ref{sec:negativity_properties}, we show that the negativity of OQ is a faithful indicator of nonclassicality.) Fig.~\ref{fig:result3d} (d) shows the negativities of the OQ and the MHQ. While the negativities of the OQ and MHQ do not coincide, they can extract exactly the same amount of work;
\begin{eqnarray}
    \langle E_f - E_i \rangle_\text{OQ} = \langle E_f - E_i \rangle_\text{MHQ},
\end{eqnarray}
as their marginal probabilities of $A$ and $B$ coincide. This signals that the negativity in a quasiprobability is necessary for increase of the extractable work, but the magnitude of negativity cannot be a faithful indicator of the amount of increase in work.

\section{conclusion}

We suggest the OQ as a work distribution in the nonequilibrium quantum thermodynamics, which can be constructed with the simple experimental settings, the EPM and the TPM scheme. The OQ satisfies the significant properties to serve as a work distribution: Marginality, TPM reproducibility, and Convex linearity. These properties allow the OQ to reproduce the JE and the average work consistent with the classical definition. Also, we show that the coherence of initial state contributes to the fluctuation, the average, and the second moment of work. For a two-level system and unbiased measurements, the negativity of OQ implies that the measurements considered are non-jointly measurable. Based on this, we show that non-joint measurability can enhance the work of the OQ beyond the classical bound, providing an operational identification of the generalized measurement incompatibility in the work extraction protocol. We further prove that, in two-dimensional systems, the OQ and the MHQ are equivalent. The equivalence reveals that the necessary and sufficient condition for the MHQ consisting of unbiased measurements to be nonnegativity is the non-joint measurability. This result suggests the potential connection between the non-joint measurability and weak value induced by the negativity of KDQ. While, in the three-level system, the negativities of the OQ and the MHQ do not coincide, they extract the same amount of work, implying that the magnitude of the negativity in a quasiprobability cannot be a faithful indicator of the amount of nonclassical work. These results highlight the significance of the coherence and non-joint measurability in the nonclassical enhancement of work. Finally, it is worth noting that the OQ framework can be implemented with relatively simple measurement settings~\cite{Kang-Da2019,Hui2025}, thereby making experimental investigations of quantum thermodynamics more feasible.

\begin{acknowledgements}
J.J. thanks Juzar Thingna and Hyukjoon Kwon for discussion. J.R. and H.R. are supported by Korea Institute of Science and Technology Information (KISTI) (No. K25L1M3C3) and by the National Research Foundation of Korea (NRF) (Grant No. RS-2023-NR119931).
\end{acknowledgements}

\setcounter{equation}{0}
\setcounter{section}{0}
\setcounter{table}{0}
\setcounter{figure}{0}
\renewcommand{\thetable}{A\arabic{table}}
\renewcommand{\d}[1]{\ensuremath{\operatorname{d}\!{#1}}}
\renewcommand{\thesection}{A\arabic{section}}
\renewcommand{\thesubsection}{A\arabic{subsection}}
\renewcommand{\theequation}{A\arabic{equation}}
\renewcommand{\thefigure}{A\arabic{figure}}

\section{Definition of OQ}
\label{sec:prop}
The distribution $q^{\text{OQ}}_{if}$ is defined based on the characteristic function of the EPM and TPM as
\begin{eqnarray}
    \chi^\text{OQ}_{0n}&=&\langle \gamma^{nf} \rangle_\text{EPM},~m=0~\text{and}~\forall n \nonumber\\
    \chi^\text{OQ}_{mn}&=&\langle \gamma^{mi}\gamma^{nf} \rangle_\text{TPM},~\forall m\neq0~\text{and}~\forall n
\end{eqnarray}
where $\gamma = \exp(2\pi j/d)$ for $j^2={-1}$ and $\langle\cdot\rangle_X$ is the expectation over the probability obtained by the respective measurement setting $X$. We assume that $\chi^\text{OQ}_{00}=1$. The OQ is associated with the inverse Fourier transformation of the characteristic functions,
\begin{eqnarray}
\label{eq:def}
    q^\text{OQ}_{if}:= \frac{1}{d^2}\sum_{m,n=0}^{d-1} \gamma^{-im-fn} \chi^\text{OQ}_{mn}.
\end{eqnarray}

By the definition~\eqref{eq:def}, we can obtain the function of OQ as
\begin{eqnarray}
    q^\text{OQ}_{if} &=& \frac{1}{d^2}\left(\sum_{m=0,n}\gamma^{-fn} \chi^\text{OQ}_{0n} + \sum_{m\neq0,n}\gamma^{-im-fn} \chi^\text{OQ}_{mn}\right) \nonumber\\
    &=&\frac{1}{d^2}\Bigg(\sum_{n,f'}\gamma^{-(f-f')n}p^\text{EPM}_{f'} \nonumber\\
    &&\qquad+\sum_{m\neq0,n,i',f'}\gamma^{-(i-i')m-(f-f')n} p^\text{TPM}_{i'f'} \Bigg) \nonumber\\
    &=&\frac{1}{d}p^\text{EPM}_f + p^\text{TPM}_{if} - \frac{1}{d}\sum_i p^\text{TPM}_{if}.
\end{eqnarray}
To obtain this result, we use $\sum_{x=0}^{d-1} \gamma^{ax} = d\delta_{a0}$.

\section{The bound of the second moment of work}
\label{sec:boundw2}
For the energies $E_f$ and $E_i$ measured at time $t_f$ and $t_i$, respectively, let the second-order moment of work $w=E_f - E_i$ be $S$ as
\begin{eqnarray}
S = \sum_{i,f}(E_f - E_i)^2 q_{if}^\text{OQ}. \nonumber
\end{eqnarray}
We can decompose $S$ into
\begin{eqnarray}
    S &=& \sum_f E_f^2 p^B_f + \sum_i E_i^2 p_i^A  - 2\sum_{i,f}E_iE_f p^\text{TPM}_{if} \\
    &&\quad -\frac{2}{d}\sum_i E_i \left(\sum_f E_f p^B_f - \sum_f E_f \sum_{i'}p^\text{TPM}_{i'f}\right), \nonumber
\end{eqnarray}
where $p^\text{TPM}_{if} = \Tr(\hat{A}_i^{1/2}\hat{\varrho}\hat{A}_i^{1/2}\hat{B}^H_f)$. We focus on the correlation term given by the TPM probability, and the Cauchy–Schwarz inequality gives its lower bound:
\begin{eqnarray}
    &&-2\Tr\left(\sum_iE_i\hat{A}_i^{1/2}\hat{\varrho}\hat{A}_i^{1/2} \sum_f E_f\hat{B}^H_f\right) \nonumber\\
    &&\quad\ge -2\left[\Tr\left(\sum_i E^2_i \hat{\varrho} \hat{A}_i\right)\right]^{1/2}\nonumber\\
    &&\qquad\times\left[\Tr\sum_i\hat{A}_i^{1/2}\hat{\varrho}\hat{A}_i^{1/2}\left(\sum_f E_f\hat{B}_f^H\right)^2\right]^{1/2} \nonumber\\
    &&\quad=-2\left(\sum_i E_i^2p^A_i\right)^{1/2}\left[\Tr\left(\hat{\varrho}_D (\hat{H}^H_f)^2\right)\right]^{1/2}.
\end{eqnarray}
Note that the measurements $A$ and $B$ are projectors. By the inequality of arithmetic and geometric means, the last expression satisfy
\begin{eqnarray}
    &&-2\left(\sum_i E_i^2p^A_i\right)^{1/2}\left[\Tr\left(\hat{\varrho}_D (\hat{H}^H_f)^2\right)\right]^{1/2} \nonumber\\
    &&\qquad \ge -\sum_i E_i^2p^A_i - \Tr\left(\hat{\varrho}_D (\hat{H}^H_f)^2\right).
\end{eqnarray}
Thus, $S$ is bounded from below:
\begin{eqnarray}
    S \ge  \Tr\left[ \hat{\varrho}_\text{off}\left(\hat{H}^H_f\right)^2\right] - \frac{2}{d}\Tr\left(\hat{H}_i\right)\Tr\left(\hat{\varrho}_\text{off}\hat{H}^H_f\right).\nonumber
\end{eqnarray}
The equality holds when $E_i \hat{P_i}^{1/2} = \hat{P}_i^{1/2}\hat{Q}$, where $\hat{P}_i=\hat{A}_i^{1/2}\hat{\varrho}\hat{A}_i^{1/2}$ and $\hat{Q}=\sum_f E_f\hat{B}^H_f$. The condition $\sum_i E_i =\Tr(\hat{H}_i)=0$ yields the following inequality:
\begin{eqnarray}
    S \ge  \Tr\left[ \hat{\varrho}_\text{off}\left(\hat{H}^H_f\right)^2\right],
\end{eqnarray}
which is equivalent to~\eqref{eq:w2bound}.\hfill\qedsymbol


\section{Proof of Lemma}
\label{sec:lemma}
Let the Bloch vectors of an input state $\hat{\varrho}$ and an unbiased measurement be $\vec{r}$ and $\vec{v}$, respectively. Then, the OQ becomes
\begin{eqnarray}
    q^\text{OQ}_{if} =\frac{1}{4}\left[1+\gamma^{x_i+x_f}\vec{v}_i\cdot\vec{v}_f+\left(\gamma^{x_i}\vec{v}_i+\gamma^{x_f}\vec{v}_f\right)\cdot\vec{r}\right].
\end{eqnarray}
This OQ function is positive semidefinite {\em if and only if } $1\pm \vec{v}_i\cdot\vec{v}_f-\Vert\vec{v}_i\pm\vec{v}_f\Vert \ge 0$, and this condition holds {\em if and only if} Busch's criterion is satisfied; $\Vert \vec{v}_i+\vec{v}_f\Vert + \Vert \vec{v}_i-\vec{v}_f \Vert \le 2$.  Rewriting the inequality by expressing $\Vert \vec{v}_i\pm\vec{v}_f\Vert \le 2 - \Vert \vec{v}_i\mp\vec{v}_f \Vert$, and squaring both sides, we obtain the positivity condition of the OQ function. \hfill\qedsymbol

\section{Modified Jarzynski equality by the OQ}
\label{sec:gammaOQ}

For the TPM scheme, the characteristic function of work $w=E_f-E_i$ is given by
\begin{eqnarray}
    \left\langle e^{-\beta w}\right\rangle_\text{TPM}&=&\sum_{i,f}p^\text{TPM}_{if}e^{-\beta(E_f-E_i)} \nonumber\\
    &=&\sum_{i,f}\Tr\left[ \hat{\varrho}_{ii}e^{\beta E_i}\hat{\Pi}_{i} e^{-\beta E_f}\Phi^\dagger_H(\hat{\Pi}_f)\right] \nonumber\\
    &=& \Tr\left[\hat{\varrho}_{D}e^{\beta \hat{H}_i} \Phi^\dagger_H(e^{-\beta \hat{H}_f})\right] \nonumber\\
    &=&\frac{Z_f}{Z_i}\Tr\left[\hat{\varrho}_{G,i}^{-1}\hat{\varrho}_{D}\Phi^\dagger_H(\hat{\varrho}_{G,f})\right]\nonumber\\
    &=&e^{-\beta \Delta F}\Gamma_\text{TPM},
\end{eqnarray}
where ${Z_f}/{Z_i}=\exp(-\beta \Delta F)$.

The OQ has additional terms determined by the probability of the EPM, $p^\text{EPM}_f$, and the marginal probability of the TPM, $\sum_i p^\text{TPM}_{if}$. The characteristic functions of the EPM and the marginal of the TPM (mTPM) are given by
\begin{eqnarray}
    \left\langle e^{-\beta w}\right\rangle_\text{EPM}&=& \frac{Z_f}{Z_i}\Tr\left(\hat{\varrho}_{G,i}^{-1}\right)\Tr\left[\hat{\varrho}\Phi^\dagger_H(\hat{\varrho}_{G,f})\right],  \nonumber\\
    \left\langle e^{-\beta w}\right\rangle_\text{mTPM}&=&\frac{Z_f}{Z_i}\Tr\left(\hat{\varrho}_{G,i}^{-1}\right)\Tr\left[\hat{\varrho}_D\Phi^\dagger_H(\hat{\varrho}_{G,f})\right].
\end{eqnarray}
Combining them, we have the characteristic function of OQ:
\begin{eqnarray}
    &&\left\langle e^{-\beta w}\right\rangle_\text{OQ} \nonumber\\
    &&\quad=\left\langle e^{-\beta w}\right\rangle_\text{TPM} + \frac{1}{d}\bigg(\left\langle e^{-\beta w}\right\rangle_\text{EPM}- \left\langle e^{-\beta w}\right\rangle_\text{mTPM}\bigg)\nonumber\\
    &&\quad=\frac{Z_f}{Z_i}\left[\Gamma_\text{TPM} + \frac{1}{d}\Tr\left(\hat{\varrho}_{G,i}^{-1}\right)\Tr\left(\hat{\varrho}_\text{off}\hat{\varrho}_{G,f}\right)\right] \nonumber\\
    &&\quad=e^{-\beta \Delta F}\Gamma_\text{OQ},
\end{eqnarray}
where $\hat{\varrho}_\text{off} = \hat{\varrho} - \hat{\varrho}_D$ denotes the off-diagonal parts of the state $\hat{\varrho}$.

\section{Difference between OQ and TPM probability}
\label{sec:traceCoh}

We have shown that the OQ can be a tool for identifying coherence through the average work and the work fluctuation~\eqref{eq:OQmod}. The capability of the OQ as an identifier of coherence also appears in its original functional form. For projective measurements, the OQ can be written as
\begin{eqnarray}
    q^\text{OQ}_{if} = p^\text{TPM}_{if} +\frac{1}{d}\Tr\left[\left(\hat{\varrho}-\hat{\varrho}_D\right)\hat{B}^H_f\right],
\end{eqnarray}
The difference between the OQ and the probability of the TPM signals the amount of coherence of the state in the basis of $A$ as
\begin{eqnarray}
\max_{B^H}\sum_{i,f}\abs{q_{if}^\text{OQ}-p_{if}^\text{TPM}} &=&\Vert\hat{\varrho}-\hat{\varrho}_D\Vert_\text{tr}. 
\end{eqnarray}
This quantity is equivalent to the $l_1$-norm coherence measure ${\cal C}_{l_1}(\hat{\varrho})$ for $d$-dimensional X-states ($d\ge2$) and qubit states~\cite{Rana2016}. By the equivalence of the OQ and the MHQ, this result implies that the difference between the MHQ and the TPM probability maximized over the measurement $B^H$ can quantify the coherence of a qubit state.

\section{Proof of Theorem $1$}
\label{sec:non-JM}

By Lemma, the OQ is positive semidefinite for all two-dimensional quantum states $\hat{\varrho}\in {\cal H}_2$ {\em if and only if} binary unbiased measurements are jointly measurable (JM). For the positive semidefinite OQ, the upper bound of extractable work is determined by ${\cal W}_\text{cl}$ derived in~\eqref{eq:bound}. The bound maximized over the sharpness of the measurements, $\max_\mu {\cal W}_\text{cl}$, is larger than or equal to ${\cal W}_\text{cl}$. This can be summarized as follows:
\begin{eqnarray}
    \text{JM} ~\Leftrightarrow~ \text{OQ}\ge0,~\forall\hat{\varrho}\in {\cal H}_2 ~\Rightarrow~ \langle w\rangle_\text{OQ} \le \max_\mu {\cal W}_\text{cl}. \nonumber
\end{eqnarray}
Thus, as the contrapositive statement, the work extraction beyond the $\max_\mu {\cal W}_\text{cl}$ implies that the measurements considered are non-jointly measurable.\hfill\qedsymbol

\section{Proof of Theorem $2$}
\label{sec:equiv}
For the measurement $A$, consider a binary measurement determined by the biasedness $x$ and the unbiasedness $\mu$ as
\begin{eqnarray}
    \hat{A}_i = \frac{1}{2}\left[\left(1+\gamma^{i} x\right)I+\vec{v}_i\cdot\vec{\sigma}\right],
\end{eqnarray}
where $\gamma=-1$, $0\le x\le 1$, and $\Vert\vec{v}_i\Vert=\mu$. The $\hat{M}_i$ is positive semidefinite when $1\pm x\pm\Vert\vec{v}_i\Vert\ge0$. This expression is the general form for representing a binary measurement defined in the two-dimensional Hilbert space~\cite{Sixia10}.

We rewrite this expression in terms of the projectors aligned to the Bloch vector of measurement $A$, $\hat{\Pi}_i=(I+\vec{v}_i/\Vert\vec{v}_i\Vert\cdot\vec{\sigma})/2$, as
\begin{eqnarray}
    \hat{A}_i = a_i\hat{\Pi}_0 + a_{i+1}\hat{\Pi}_1.
\end{eqnarray}
where $a_i=[1+\gamma^i(x+\mu)]/2\ge 0  $ and $a_i+a_{i+1} = 1$. As $\hat{\Pi}_i\hat{\Pi}_j=\delta_{ij}\hat{\Pi}_i$, we can obtain the square root of the measurement by taking square root at each coefficient as $\hat{A}^{1/2}_i= a^{1/2}_i\hat{\Pi}_0 + a^{1/2}_{i+1}\hat{\Pi}_1$.

The equivalence of the OQ and MHQ is based on the following identity:
\begin{eqnarray}
    &&\sum_i \hat{A}^{1/2}_i\hat{B}^H_f\hat{A}^{1/2}_i\nonumber\\
    &&=2\hat{A}^{1/2}_i\hat{B}^H_f\hat{A}^{1/2}_i  + \hat{A}^{1/2}_{i+1}\hat{B}^H_f\hat{A}^{1/2}_{i+1}-\hat{A}^{1/2}_i\hat{B}^H_f\hat{A}^{1/2}_i\nonumber\\
    &&= 2\hat{A}^{1/2}_i\hat{B}^H_f\hat{A}^{1/2}_i + a_{i+1}\hat{\Pi}_i\hat{B}^H_f\hat{\Pi}_i + a_i\hat{\Pi}_{i+1}\hat{B}^H_f\hat{\Pi}_{i+1}\nonumber\\
    &&\quad -a_i\hat{\Pi}_i\hat{B}^H_f\hat{\Pi}_i - a_{i+1}\hat{\Pi}_{i+1}\hat{B}^H_f\hat{\Pi}_{i+1}  \nonumber\\
    &&= 2\hat{A}^{1/2}_i\hat{B}^H_f\hat{A}^{1/2}_i +a_{i+1}(I-\hat{\Pi}_{i+1})\hat{B}^H_f(I-\hat{\Pi}_{i+1})\nonumber\\
    &&\quad+a_i(I-\hat{\Pi}_i)\hat{B}^H_f(I-\hat{\Pi}_i) -a_i\hat{\Pi}_i\hat{B}^H_f\hat{\Pi}_i - a_{i+1}\hat{\Pi}_{i+1}\hat{M}_f\hat{\Pi}_{i+1} \nonumber \\
    &&= 2\hat{A}^{1/2}_i\hat{B}^H_f\hat{A}^{1/2}_i + \hat{B}^H_f - \hat{A}_i\hat{B}^H_f - \hat{B}^H_f\hat{A}_i.
\end{eqnarray}
We arrange this identity as
\begin{eqnarray}
    &&\hat{A}^{1/2}_i\hat{B}^H_f\hat{A}^{1/2}_i + \frac{1}{2}\left( \hat{B}^H_f - \sum_i \hat{A}^{1/2}_i\hat{B}^H_f\hat{A}^{1/2}_i \right)\nonumber\\
    &&\quad=\frac{1}{2}\left(\hat{A}_i\hat{B}^H_f+\hat{B}^H_f\hat{A}_i\right),~\forall i,f,
\end{eqnarray}
and multiplying both sides by a quantum state $\hat{\varrho}$ and taking a trace lead to the equivalence of the MHQ and the OQ. \hfill\qedsymbol

\section{Properties of the OQ negativity}
\label{sec:negativity_properties}

Like the KDQ negativity~\cite{Lostaglio2023kdq}, the negativity of OQ has some useful properties to witness nonclassicality:

\begin{enumerate}[label={(N\arabic*)}]
    \item {\em Faithfulness}: ${\cal N}\left[q^\text{OQ}\right]=0$ if and only if $q^\text{OQ}$ is a probability distribution.

    \item {\em Non-commutativity witness}: If ${\cal N}(q^\text{OQ})\ge 0$, $[\hat{\varrho},\hat{A}_i]\neq0$ and $[\hat{A}_i,\hat{B}^H_f]\neq0$ for some indices $i$ and $f$.
    
    \item {\em Convexity}: For $\sum_k p_k q^\text{OQ}_k(\hat{\varrho})$ where $p_k\ge0$ $\forall k$ and $\sum_k p_k = 1$, ${\cal N}[\sum_k p_k q^\text{OQ}_k(\hat{\varrho})]\le\sum_k p_k {\cal N}[q_k^\text{OQ}(\hat{\varrho})]$.
    
    \item {\em Monotone under decoherence}: For the decoherence process ${\cal E}_D[\cdot] = (1-s)I+sD[\cdot]$ where $s\in[0,1]$ and $D[\cdot]$ is a transformation which remove off-diagonal elements either in the basis of the measurement $A$ or $B$, ${\cal N}[q^\text{OQ}({\cal E}_D[\hat{\varrho}])] \le {\cal N}[q^\text{OQ}(\hat{\varrho})]$. 
    
    \item {\em Monotone under coarse-graining}: For $\tilde{q}^\text{OQ}_{IF}(\hat{\varrho}) := \sum_{i\in I, f \in F}q^\text{OQ}_{if}(\hat{\varrho})$, where $I$ and $F$ are disjoint subsets that partition the indices $\{i\}$ and $\{f\}$, respectively, then ${\cal N}\left[ \tilde{q}^\text{OQ}(\hat{\varrho}) \right] \le {\cal N}\left[q^\text{OQ}(\hat{\varrho})\right]$.

\end{enumerate}

The following presents the proofs of (N$1$)--(N$5$):

{Proof of (N$1$)}.---If any element of the OQ is negative, for some $i$ and $f$, $\vert{q^\text{OQ}_{if}}\vert>q^\text{OQ}_{if}$. This leads to $\sum_{if} (\vert{q^\text{OQ}_{if}}\vert - q^\text{OQ}_{if}) >0$, which is equivalent to ${\cal N}[q^\text{OQ}]>0$ as $\sum_{if}q^\text{OQ}_{if}=1$. The converse is also true.\hfill\qed

{Proof of (N$2$)}.---If $[\hat{\varrho},\hat{A}_i]=0$ or $[\hat{A}_i, \hat{B}^H_f]=0$ $\forall i,j$, the probability of TPM can be written $p^\text{TPM}_{if} = \Tr\left(\hat{\varrho}\hat{A}_i\hat{B}^H_f\right)$. It follows that the marginal of the TPM probability and the EPM are the same $p^\text{EPM}_f = \sum_f p^\text{TPM}_{if}$. In this condition, the OQ function is positive semidefinite as $q^\text{OQ}_{if} = p^\text{TPM}_{if}$ $\forall i,f$. The contrapositive is (N$2$).\hfill\qed

{Proof of (N$3$)}.---By the convexity of the absolute function, $\sum_{if} \vert{pq^\text{OQ}_{if,1} + (1-p)q^\text{OQ}_{if,2}}\vert -1 \le p(\sum_{if}\vert{q^\text{OQ}_{if,1}}\vert-1) + (1-p)(\sum_{if}\vert{q^\text{OQ}_{if,2}}\vert-1)$ holds. Thus, the negativity of the OQ satisfies the convexity.\hfill\qed

{Proof of (N$4$)}.---By the property of convexity (N$3$), ${\cal N}[q^\text{OQ}({\cal E}_D(\hat{\varrho}))] \le (1-s){\cal N}(q^\text{OQ}(\hat{\varrho})) + s {\cal N}(q^\text{OQ}(D[\hat{\varrho}]))$. As $D[\hat{\varrho}]$ commutes either with $A$ or $B$, the OQ of $D[\hat{\varrho}]$ is positive semidefinite by the property (N$2$). Thus, ${\cal N}[q^\text{OQ}({\cal E}_D(\hat{\varrho}))] \le (1-s){\cal N}(q^\text{OQ}(\hat{\varrho})) \le {\cal N}(q^\text{OQ}(\hat{\varrho}))$.\hfill\qed

{Proof of (N$5$)}.---${\cal N}[\tilde{q}^\text{OQ}_{IF}]=\sum_{I,J}\vert{\sum_{i\in I, f\in F}q^\text{OQ}_{if}}\vert-1$ and this term is upper bounded by the convexity (N$3$) as $\sum_{I,J}\vert{\sum_{i\in I, f\in F}q^\text{OQ}_{if}}\vert-1\le \sum_{I,J}\sum_{i\in I, f\in F}\vert{q^\text{OQ}_{if}}\vert-1$. Thus, ${\cal N}[\tilde{q}^\text{OQ}(\hat{\varrho})] \le {\cal N}[q^\text{OQ}(\hat{\varrho})]$. \hfill\qed


%

\end{document}